\documentclass[12pt]{article}
\usepackage{amsfonts}

\usepackage{mathrsfs}
\usepackage{amscd}
\usepackage{amsmath,amsfonts,amssymb,amscd}
\usepackage{indentfirst,graphics,epsfig,psfrag}
\input{epsf}

\usepackage{amsmath,amssymb,amsthm, amscd, amsfonts, graphicx,ifpdf}
\usepackage{lineno}

\newtheorem{thm}{Theorem}[section]

\newtheorem{cor}{Corollary}[section]

\newtheorem{lem}{Lemma}[section]

\def\qed{\nopagebreak\hfill{\rule{4pt}{7pt}}
\medbreak}

\def\pf{\noindent {\it Proof.} }

\title{\bf Note on the complexity of\\ deciding the rainbow connectedness\\ for
bipartite graphs\footnote{Supported by NSFC and the Fundamental
Research Funds for the Central Universities. }}

\author{
\small Shasha Li, Xueliang Li\\
\small Center for Combinatorics and LPMC-TJKLC\\
\small Nankai University, Tianjin 300071, China.\\
\small  Email: lss@cfc.nankai.edu.cn, lxl@nankai.edu.cn\\
}
\date{}
\begin{document}

\maketitle

\begin{abstract}
A path in an edge-colored graph is said to be a rainbow path if no
two edges on the path have the same color. An edge-colored graph is
(strongly) rainbow connected if there exists a rainbow (geodesic)
path between every pair of vertices. The (strong) rainbow connection
number of $G$, denoted by ($scr(G)$, respectively) $rc(G)$, is the
smallest number of colors that are needed in order to make $G$
(strongly) rainbow connected. Though for a general graph $G$ it is
NP-Complete to decide whether $rc(G)=2$ , in this paper, we show
that the problem becomes easy when $G$ is a bipartite graph.
Moreover, it is known that deciding whether a given edge-colored
(with an unbound number of colors) graph is rainbow connected is
NP-Complete. We will prove that it is still NP-Complete even when
the edge-colored graph is bipartite. We also show that a few NP-hard
problems on rainbow connection are indeed NP-Complete.
\\[3mm]
\end{abstract}

\section{Introduction}

We follow the terminology and notations of \cite{Bondy} and all
graphs considered here are always finite and simple.

Let $G$ be a nontrivial connected graph on which is defined a
coloring $c$: $E(G)\rightarrow \{1,2,\ldots,k\}$, $k\in \mathbb{N}$,
of the edges of $G$, where adjacent edges may be colored the same. A
$u-v$ path $P$ in $G$ is a $rainbow\ path$ if no two edges of $P$
are colored the same. The graph $G$ is $rainbow\ connected$ $(with\
respect\ to\ c)$ if $G$ contains a rainbow $u-v$ path for every two
vertices $u$ and $v$ of $G$. In this case, the coloring $c$ is
called a $rainbow\ coloring$ of $G$. If $k$ colors are used, then
$c$ is a $rainbow$ $k$-$coloring$. The $rainbow\ connection\ number$
of $G$, denoted by $rc(G)$, is the smallest number of colors that
are needed in order to make $G$ rainbow connected. A $rainbow\ u-v\
geodesic$ in $G$ is a $rainbow\ u-v\ path$ of length $d(u,v)$, where
$d(u,v)$ is the distance between $u$ and $v$. The graph $G$ is
$strongly\ rainbow\ connected $ if there exists a rainbow $u-v$
geodesic for any two vertices $u$ and $v$ in $G$. In this case, the
coloring $c$ is called a $strong\ rainbow\ coloring$ of $G$.
Similarly, we define the $strong\ rainbow\ connection\ number$ of a
connected graph $G$, denoted by $src(G)$, as the smallest number of
colors that are needed in order to make $G$ strong rainbow
connected. Clearly, we have $diam(G)\leq rc(G)\leq scr(G)\leq m$,
where $diam(G)$ denotes the diameter of $G$ and $m$ is the number of
edges of $G$. Moreover, it is easy to verify that $src(G)=rc(G)=1$
if and only if $G$ is a complete graph, that $rc(G)=2$ if and only
if $src(G)=2$, and that $rc(G)=n-1$ if and only if $G$ is a tree.
The concepts of rainbow connectivity and strong rainbow connectivity
were first introduced by Chartrand et al. in \cite{Chartrand} as a
means of strengthening the connectivity. Subsequent to this paper,
the problem has received attention by several people and the
complexity as well as upper bounds for the rainbow connection number
have been studied.

In \cite{Caro}, Caro et al. conjectured that computing $rc(G)$ is an
NP-Hard problem, as well as that even deciding whether a graph has
$rc(G)=2$ is NP-Complete. In \cite{Chakraborty}, Chakraborty et al.
confirmed this conjecture. In \cite{Prabhanjan}, the complexity of
computing $rc(G)$ and $src(G)$ was studied further. It was shown
that given any natural number $k\geq 3$ and a graph $G$, it is
NP-hard to determine whether $rc(G)\leq k$. Moreover, for $src(G)$,
it was shown that given any natural number $k\geq 3$ and a graph
$G$, determining whether $src(G)\leq k$ is NP-hard even when $G$ is
bipartite. In this paper, we will point out that the problems in
\cite{Prabhanjan} are, in fact, NP-Complete. Though for a general
graph $G$ it is NP-Complete to decide whether $rc(G)=2$
\cite{Chakraborty}, we show that the problem becomes easy when $G$
is a bipartite graph. Moreover, it is NP-Complete to decide whether
a given edge-colored (with an unbound number of colors) graph is
rainbow connected \cite{Chakraborty}. We will prove that it is still
NP-Complete even when the edge-colored graph is bipartite.

\section{Main results}

At first, we restate several results in \cite{Chakraborty} and
\cite{Prabhanjan}.

\begin{lem}\label{lem1}(\cite{Chakraborty})
Given a graph $G$, deciding if $rc(G)=2$ is NP-Complete. In
particular, computing $rc(G)$ is NP-Hard.
\end{lem}

\begin{lem}\label{lem2}(\cite{Prabhanjan})
For every $k\geq 3$, deciding whether $rc(G)\leq k$ is NP-Hard.
\end{lem}

\begin{lem}\label{lem3}(\cite{Prabhanjan})
Deciding whether the rainbow connection number of a graph is at most
$3$ is NP-Hard even when the graph $G$ is bipartite.
\end{lem}

\begin{lem}\label{lem4}(\cite{Prabhanjan})
For every $k\geq 3$, deciding whether $src(G)\leq k$ is NP-Hard even
when $G$ is bipartite.
\end{lem}

We will show that ``NP-hard" in the above results can be replaced by
``NP-Complete" if $k$ is any fixed integer. It suffices to show that
these problems belong to the class NP for any fixed $k$. In fact,
from the proofs in \cite{Prabhanjan}, for the problems in Lemmas
\ref{lem2} and \ref{lem4}, ``For every $k\geq 3$" can be replaced by
``For any fixed $k\geq 3$".
\begin{thm}\label{thm1}
For any fixed $k\geq 2$, given a graph $G$, deciding whether
$rc(G)\leq k$ is NP-Complete.
\end{thm}
\pf By Lemmas \ref{lem1} and \ref{lem2}, it will suffice to show
that the problem in Lemma \ref{lem2} is in NP. Therefore, if given
any instance of the problem whose answer is `yes', namely a graph
$G$ with $rc(G)\leq k$, we want to show that there is a certificate
validating this fact which can be checked in polynomial time.

Obviously, a rainbow $k$-coloring of $G$ means that $rc(G)\leq k$.
For checking a rainbow $k$-coloring, we need only check whether $k$
colors are used and for any two vertices $u$ and $v$ of $G$, whether
there exists a rainbow $u-v$ path. Notice that for two vertices
$u,v$, there are at most $n^{l-1}$ $u-v$ paths of length $l$, since
if let $P=ut_{1}t_{2}\cdots t_{l-1}v$, there are less than $n$
choices for each $t_{i}$ $(i\in \{1,2,\ldots,l-1\})$. Therefore, $G$
contains at most $\Sigma_{l=1}^{k} n^{l-1}\leq kn^{k-1}\leq n^{k}$
$u-v$ paths of length no more than $k$. Then check these paths in
turn until find one path whose edges have distinct colors or no such
paths at all. It follows that the time used for checking is at most
$O(n^{k}\cdot n\cdot n^{2})=O(n^{k+3})$. Since $k$ is a fixed
integer, we conclude that the certificate, namely a rainbow
$k$-coloring of $G$, can be checked in polynomial time. The proof is
now complete.\qed

The next theorem can be obtained similarly.

\begin{thm}\label{thm2}
For any fixed $k\geq 2$, given a graph $G$, deciding whether
$src(G)\leq k$ is NP-Complete.
\end{thm}
\pf Since $rc(G)=2$ if and only if $src(G)=2$, by Lemmas \ref{lem1}
and \ref{lem4}, it will suffice to show that the problem in Lemma
\ref{lem4} is in NP.

From the proof of Theorem \ref{thm1}, it is clear that for any two
vertices $u$ and $v$ of $G$, the existence of a $u-v$ path of length
$l$ $(\leq k)$ can be decided in time $O(n^{l-1})$. Therefore, if we
check each integer $l\leq k$ in turn, we can either find an integer
$l$ such that there is a $u-v$ path of length $l$ but no $u-v$ path
of length less than $l$, or conclude that there is no $u-v$ path of
length at most $k$. In the former case, the integer $l$ is exactly
the distance $d(u,v)$ between $u$ and $v$ and then check the colors
of edges of each $u-v$ path of length $d(u,v)$ in turn. Similarly to
the proof of Theorem \ref{thm1}, we can obtain that the certificate,
namely a strong rainbow $k$-coloring of $G$, can be checked in
polynomial time. The proof is complete.\qed

We know that given a graph $G$, deciding if $rc(G)=2$ is
NP-Complete. Surprisingly, if $G$ is a bipartite graph, the problem
turns out to be easy. Before giving the proof, we first introduce
the following result of \cite{Chartrand}.

\begin{lem}\label{lem5}(\cite{Chartrand})
For integers $s$ and $t$ with $2\leq s\leq t$,
$$
rc(K_{s,t})=min\{\lceil\sqrt[s]{t}\ \rceil,\ 4\}.
$$
\end{lem}

\begin{thm}\label{thm3}
For a bipartite graph $G$, deciding whether $rc(G)=2$ can be solved
in polynomial time.
\end{thm}
\pf Obviously if $G$ is not a complete bipartite graph, there must
exist two nonadjacent vertices $x$ and $y$ in the different parts of
$G$. But then the distance $d(x,y)$ must be at least $3$. We know
that $d(x,y)\leq diam(G)\leq rc(G)$. It follows that $rc(G)\neq 2$.
Therefore, only when $G$ is a complete bipartite graph $K_{s,t}$
$(s\leq t)$, it is possible that $rc(G)=2$. If $s=1$, then $G$ is a
star and $rc(G)=t$. Otherwise by Lemma \ref{lem5},
$rc(G)=min\{\lceil\sqrt[s]{t}\ \rceil,\ 4\}$. One needs only to
check if $1<t\leq 2^s$, which can be done by simple computation and
comparison. Moreover, it is clear that checking whether $G$ is a
complete bipartite graph can be done in polynomial time. The proof
is complete.\qed

Then by Lemma \ref{lem3}, the following result is immediate.

\begin{cor}\label{cor1}
Given a bipartite graph $G$, deciding if $rc(G)=3$ is NP-Complete.
\end{cor}

As shown in the proof of Theorem \ref{thm1}, given an edge-coloring
of a graph, if the number of colors is constant, then we can verify
whether the colored graph is rainbow connected in polynomial time.
However, in \cite{Chakraborty}, chakraborty et al. showed that if
the coloring is arbitrary, the problem becomes NP-Complete.

\begin{lem}\label{lem6}(\cite{Chakraborty})
The following problem is NP-Complete: Given an edge-colored graph
$G$, check whether the given coloring makes $G$ rainbow connected.
\end{lem}

Now we prove that even when $G$ is bipartite, the problem is still
NP-Complete.

\begin{thm}\label{thm4}
Given an edge-colored bipartite graph $G$, checking whether the
given coloring makes $G$ rainbow connected is NP-Complete.
\end{thm}
\pf By Lemma \ref{lem6}, it will suffice by showing a polynomial
reduction from the problem in Lemma \ref{lem6}.

Given a graph $G=(V,E)$ and an edge-coloring $c$ of $G$, we will
construct an edge-colored bipartite graph $G'$ such that $G$ is
rainbow connected if and only if $G'$ is rainbow connected.

Now for each edge $e\in E(G)$, subdivide $e$ by a new vertex
$v_{e}$. The obtained graph is exactly $G'$ and $(X,Y)$ is a
bipartition of $G'$, where $X=V(G)$ and $Y=\{v_{e}\ |\ e\in E(G)\}$.
Then the edge-coloring $c'$ of $G'$ is defined by for each edge
$e=v_{i}v_{j}\in E(G)$ $(i\leq j)$, $c'(v_{i}v_{e})=c(e)$ and
$c'(v_{j}v_{e})=l_{e}$, where $l_{e}$ is a new color and different
from the colors used in $c$ and if $e\neq e'$, then $l_{e}\neq
l_{e'}$.

If $c'$ is a rainbow coloring of $G'$, then any two vertices $u$ and
$v$ are connected by a rainbow path $P'_{u,v}$, including every pair
of vertices in $X=V(G)$. Clearly, by contracting edges which are
assigned new colors, $P'_{u,v}$ can be converted to a rainbow path
$P_{u,v}$ of $G$ (with respect to $c$), where $u,v\in V(G)$. It
follows that the coloring $c$ makes $G$ rainbow connected.

To prove the other direction, assume that for every two vertices
$v_{t}$ and $v_{t'}$ of $G$, there always exists a rainbow path
$P_{v_{t}v_{t'}}=v_{t}v_{t_1}v_{t_2}\ldots v_{t'}$. Now for each
pair $(v_{t},v_{t'})$ of vertices in $V(G')$, if $v_{t},v_{t'}\in
X=V(G)$, then
$P'_{v_{t}v_{t'}}=v_{t}v_{e_{m_1}}v_{t_1}v_{e_{m_2}}v_{t_2}\ldots
v_{e_{m_j}} v_{t'}$ is a rainbow path in $G'$, where the vertex
$v_{e_{m_i}}$ subdivides the edge $e_{m_i}=v_{t_{i-1}}v_{t_{i}}$ of
$G$, $i\in\{1,\ldots,j\}$ (when $i=1$, the edge is $v_{t}v_{t_{1}}$
and when $i=j$, the edge is $v_{t_{j-1}}v_{t'}$). If
$v_{t},v_{t'}\in Y$, then there exist two edges
$e_{1}=v_{i_{1}}v_{j_{1}}$ and $e_{2}=v_{i_{2}}v_{j_{2}}$
($i_{1}\leq j_{1}$ and $i_{2}\leq j_{2}$) such that in $G'$ $v_{t}$
and $v_{t'}$ subdivide $e_{1}$ and $e_{2}$, respectively. Since
$v_{j_{1}},v_{j_{2}}\in X=V(G)$, we can find a rainbow path
$P'_{v_{j_{1}}v_{j_{2}}}$ in $G'$ which can be converted to a
rainbow $v_{t}-v_{t'}$ path
$P'_{v_{t}v_{t'}}=v_{t}v_{j_{1}}P'_{v_{j_{1}}v_{j_{2}}}v_{j_{2}}v_{t'}$.
The proof of the case that $v_{t}\in X$ and $v_{t'}\in Y$ is
similar. Therefore $G'$ is rainbow connected with respect to $c'$.
The proof is complete.\qed

\end{document}